\documentclass[twocolumn, showpacs, nofootinbib, aps, prd]{revtex4}

\usepackage{graphicx}
\usepackage{dcolumn}
\usepackage{bm}
\usepackage{hyperref}

\begin{document}

\title{Information on the Pion Distribution Amplitude from the Pion-Photon Transition Form Factor with the Belle and BaBar Data}

\author{Xing-Gang Wu}
\email[email: ]{wuxg@cqu.edu.cn}
\affiliation{Department of Physics, Chongqing University, Chongqing 400044, P.R. China}

\author{Tao Huang}
\email[email: ]{huangtao@ihep.ac.cn}
\author{Tao Zhong}
\email[email: ]{zhongtao1980@cqu.edu.cn}
\affiliation{Institute of High Energy Physics and Theoretical Physics Enter for Science Facilities, Chinese Academy of Sciences, Beijing 100049, P.R. China}

\date{\today}

\begin{abstract}

The pion-photon transition form factor (TFF) provides strong constraints on the pion distribution amplitude (DA). We perform an analysis of all existing data (CELLO, CLEO, BaBar, Belle) on the pion-photon TFF by means of light-cone pQCD approach in which we include the next-to-leading order correction to the valence-quark contribution and estimate the non-valence-quark contribution by a phenomenological model based on the TFF's limiting behavior at both $Q^2\to 0$ and $Q^2\to\infty$. At present, the pion DA is not definitely determined, it is helpful to have a pion DA model that can mimic all the suggested behaviors, especially to agree with the constraints from the pion-photon TFF in whole measured region within a consistent way. For the purpose, we adopt the conventional model for pion wavefunction/DA that has been constructed in our previous paper \cite{hw1}, whose broadness is controlled by a parameter $B$. We fix the DA parameters by using the CELLO, CLEO, BABAR and Belle data within the smaller $Q^2$ region ($Q^2 \leq 15$ GeV$^2$), where all the data are consistent with each other. And then the pion-photon TFF is extrapolated into larger $Q^2$ region. We observe that the BABAR favors $B=0.60$ which has the behavior close to the Chernyak-Zhitnitsky DA, whereas the recent Belle favors $B=0.00$ which is close to the asymptotic DA. We need more accurate data at large $Q^2$ region to determine the precise value of $B$, and the definite behavior of pion DA can be concluded finally by the consistent data in the coming future.

\end{abstract}


\maketitle

The pion-photon transition form factor (TFF), $F_{\pi \gamma}(Q^2)$, which relates two photons with one lightest meson, provides a good platform to study the property of pion distribution amplitude (DA). Because higher helicity and higher twist structures give negligible contributions to the pion-photon TFF~\cite{higher1,higher2}, one can extract useful information on the shape of the leading-twist pion DA by comparing the estimated result of $F_{\pi \gamma}(Q^2)$ with the measured one.

Experimentally, the pion-photon TFF is determined by measuring the process $e^+ e^-\to e^+ e^-\pi^0$ in the single-tag mode, where one of the outing electron (tagged) is detected while the other electron (untagged) is scattered at a small angle. The tagged electron emits a highly off-shell photon with momentum transfer $Q^2$ and the momentum transfer to the untagged electron is near zero. The pion-photon TFF has first been measured by the CELLO collaboration with $Q^2<3$ GeV$^2$~\cite{CELLO}. Later on, the CLEO collaboration measured such form factor with a broader range of $Q^2\in$ $[1.5,9.2]$~GeV$^2$~\cite{CLEO}, and the BABAR collaboration measured the form factor with $Q^2\in$ $[4,40]$~GeV$^2$~\cite{babar}. The newly released data by the Belle Collaboration~\cite{Belle}, seem to be dramatically different from those reported by the BABAR Collaboration~\cite{babar}. Instead of a pronounced growth of the TFF at high $Q^2$ region, observed by BABAR, the Belle data are compatible with the well-known asymptotic prediction~\cite{lb}, i.e. $Q^2 F_{\pi \gamma} (Q^2)$ tends to be a constant ($2f_\pi$) for asymptotic DA $\phi^{as}_\pi(x,Q^2)|_{Q^2\rightarrow \infty}=6x(1-x)$. Here the pion decay constant $f_{\pi}=92.4 \pm 0.25$~MeV~\cite{pdg}.

At present, there is still no definite conclusion on whether pion DA is in
asymptotic-like form~\cite{lb} or in Chernyak-Zhitnitsky (CZ)-like form~\cite{cz}, or in a flat-like form~\cite{flat}. It would be helpful to have a consistent pion DA model that can mimic all these behaviors and can explain the pion-photon TFF data in a more consistent way \footnote{The conventional Gegenbauer form for pion DA can not be directly adopted for such purpose, since as shown by a next-to-next-leading-order (NNLO) calculation in Ref.\cite{nlotff}, even using the optimal Brodsky-Lepage-Mackenzie (BLM) renormalization scale~\cite{blm} (or its improved version: Principle-Maximum-Conformality scale~\cite{pmc}) up to next-to-leading-order (NLO), the DA with big second Gegenbauer moments, such as the CZ-DA, can not explain the pion-photon TFF in small $Q^2$ region.}. By comparing their estimates of the pion-photon TTF within whole $Q^2$ region, one will obtain useful information/constraint on the pion DA. This is the main purpose of the present paper.

Generally, the pion-photon TFF can be divided into two parts,
\begin{equation}
F_{\pi \gamma}(Q^2)=F^{(V)}_{\pi\gamma}(Q^2)+ F^{(NV)}_{\pi\gamma}(Q^2) ,
\end{equation}
where $F^{(V)}_{\pi\gamma}(Q^2)$ stands for the usual valence-quark part, $F^{(NV)}_{\pi\gamma}(Q^2)$ is the non-valence-quark part that is related to the higher Fock-states of pion. Usually, $F^{(NV)}_{\pi \gamma}(Q^2)$ will be suppressed by at least $1/Q^2$ to $F^{(V)}_{\pi \gamma}(Q^2)$ in the limit $Q^2\to\infty$. Then at large $Q^2$ region, the non-valence Fock-state part $F^{(NV)}_{\pi \gamma}(Q^2)$ is negligible. However, it will give sizable contribution at small $Q^2$ region and should be kept for a sound estimation.

The valence-quark contribution $F^{(V)}_{\pi\gamma}(Q^2)$ dominates only as $Q^2$ becomes very large. Under the light-cone pQCD approach~\cite{lb}, and by keeping the $k_\bot$-corrections in both the hard-scattering amplitude and the WF, $F^{(V)}_{\pi \gamma}(Q^2)$ has been calculated up to NLO~\cite{higher2,hw1,hw2,bhl,hnli}, after further doing the integration over the azimuth angle, we obtain,
\begin{widetext}
\begin{eqnarray}
F^{(V)}_{\pi \gamma}(Q^2)&=& \frac{1}{4\sqrt{3}\pi^2}\int_0^1\int_0^{x^2 Q^2}\frac{dx}{x Q^2}\left[1-\frac{\alpha_s(Q^2)}{3\pi}\left(\ln\frac{Q^2}{xQ^2+k_\perp^2} +2\ln{x}+3- \frac{\pi^2}{3} \right)\right] \Psi_{q\bar{q}}(x,k_\perp^2) d k^2_\perp , \label{ffv}
\end{eqnarray}
\end{widetext}
where $k_\perp=|\mathbf{k}_\perp|$. Here, without loss of generality, the usual assumption that the pion WF depending on $\mathbf{k}_\perp$ through $k_\perp^2$ only has been implicitly adopted.

The non-valence-quark contribution $F^{(NV)}_{\pi \gamma}(Q^2)$ can be estimated by the phenomenological model~\cite{hw1,hw2}:
\begin{equation}\label{ffnv}
F^{(NV)}_{\pi \gamma}(Q^2)=\frac{\alpha}{(1+Q^2/\kappa^2)^2} .
\end{equation}
The parameters $\kappa=\sqrt{-\frac{F_{\pi\gamma}(0)} {\frac{\partial}{\partial Q^2}F^{(NV)}_{\pi \gamma}(Q^2)|_{Q^2\to 0}}}$ and $\alpha=\frac{1}{2}F_{\pi\gamma}(0)$ are determined by the limiting behavior of $F^{(NV)}_{\pi \gamma}(Q^2)$ at $Q^2\to 0$; i.e. two limiting behavior of $F^{(NV)}_{\pi \gamma}(Q^2)$ at $Q^2\to 0$ can be written as
\begin{equation}
F^{(NV)}_{\pi \gamma}(0) =F^{(V)}_{\pi \gamma}(0) =\frac{1}{8\sqrt{3}\pi^2}\int dx\Psi_{q\bar{q}}(x,\mathbf{0}_\perp) ,
\end{equation}
and
\begin{eqnarray}
&&\frac{\partial}{\partial Q^2}F^{(NV)}_{\pi \gamma}(Q^2)|_{Q^2\to 0} \nonumber\\
&=& \frac{1}{8\sqrt{3}\pi^2} \left[\frac{\partial}{\partial Q^2}\int_0^1\int_{0}^{x^2 Q^2}\left(\frac{\Psi_{q\bar{q}}(x,k_\perp^2)}{x^2 Q^2}\right)dx
dk_\perp^2\right]_{Q^2\to 0} ,
\end{eqnarray}
where $x'=1-x$.

Eqs.(\ref{ffv},\ref{ffnv}) show that the pion-photon TFF depends on how well we know the pion wavefunction (WF). Inversely, if we know pion-photon TFF well either theoretically or experimentally, we can determine what the pion WF and hence its DA will like.

Following the idea of Refs.\cite{spin1,spin2,spin3}, the authors of Refs.\cite{hw1,hw2} have constructed a pion WF $\Psi_{q\bar{q}}(x,\mathbf{k}_\perp)$ with the help of the BHL prescription~\cite{bhl} and the Melosh rotation~\cite{melosh}; i.e. the full form of the pion WF can be written as
\begin{equation}\label{wave}
\Psi_{q\bar{q}}(x,{\bf k}_{\perp})=\sum_{\lambda_{1}\lambda_{2}} \chi^{\lambda_{1} \lambda_{2}}(x,{\bf k}_{\perp}) \Psi^{R}_{q\bar{q}}(x,{\bf k}_{\perp}) ,
\end{equation}
with the spatial WF
\begin{equation}
\Psi^{R}_{q\bar{q}}(x,{\bf k}_{\perp})= A \varphi_\pi(x) \exp\left[-\frac{{\bf k}_{\perp}^2 +m_q^2}{8{\beta}^2x(1-x)}\right] .
\end{equation}
Here $\varphi_\pi(x)\neq 1$ denotes the deviation from the asymptotic form, which can be expanded in Gegenbauer polynomials, and by keeping its first two terms, we obtain
\begin{equation}
\varphi_\pi(x)=1+B\times C^{3/2}_2(2x-1) .
\end{equation}
The indexes $\lambda_1$ and $\lambda_2$ are helicity states of the two constitute quarks, $\chi^{\lambda_{1}\lambda_{2}}(x,{\bf k}_{\perp})$ stands for the spin-space WF coming from the Wigner-Melosh rotation. The spin-space WF $\chi^{\lambda_{1}\lambda_{2}}(x,{\bf k}_{\perp})$ can be found in Refs.\cite{spin1,spin2,spin3}. The parameter $m_q$ stands for the light constitute-quark mass. The normalization constant $A$, the harmonic scale $\beta$ and the light constitute-quark mass $m_q$ are constrained by some reasonable constraints, such as its normalization condition, the constraint derived from $\pi^0\rightarrow \gamma\gamma$ decay amplitude~\cite{bhl}, the reasonable values for the probability $P_{q\bar{q}}$ and the squared charged mean radius $\langle r^2_{\pi^+}\rangle^{q\bar{q}}$ of the valence quark state.

\begin{table}[ht]
\caption{Pion DA parameters for $m_q=0.30$ GeV, and its probability $P_{q\bar{q}}$, charged mean radius $\sqrt{\langle r^2_{\pi^+}\rangle^{q\bar{q}}}$ (unit: $fm$) and the second Gegenbauer moment $a_2(\mu_0^2)$. }
\begin{center}
\begin{tabular}{|c||c|c||c|c|c|c|}
\hline\hline ~~~$B$~~~ & ~$A ({\rm GeV}^{-1})$~& ~$\beta ({\rm GeV})$~& ~$P_{q\bar{q}}$~& ~$\sqrt{\langle r^2_{\pi^+}\rangle^{q\bar{q}}}$~ & ~$a_2(\mu_0^2)$~ \\
\hline
~$0.00$~ & ~$25.06$~& ~$0.586$~& ~$63.5\%$~& ~$0.341$~ & ~$0.03$~ \\
\hline
~$0.30$~ & ~$20.26$~& ~$0.668$~& ~$62.0\%$~& ~$0.378$~ & ~$0.36$~ \\
\hline
~$0.60$~ & ~$16.62$~& ~$0.745$~& ~$79.9\%$~& ~$0.451$~ & ~$0.68$~ \\
\hline\hline
\end{tabular}
\label{tab1}
\end{center}
\end{table}

Moreover, it is found that the present experimental data such as CELLO, CLEO, BABAR and Belle data are consistent with each other within smaller $Q^2$ region ($Q^2 \leq 15$ GeV$^2$), so we can use these TFF data in small $Q^2$-region for further constraining the WF parameters~\cite{higher1,constrain}. In fact, we think only in this way can one obtain a consistent pion-photon TFF within the whole $Q^2$ region. As a useful reference, we present the typical parameters for $m_q=0.30$ GeV in Table~\ref{tab1}.

As argued in Ref.\cite{bhl}, the leading Fock-state contributes to $F_{\pi\gamma}(0)$ only half and the remaining half should be come from the higher Fock-states as $Q^2\to 0$. And then both contributions from the leading Fock-state and the higher Fock-states are needed to get the correct $\pi^0\to\gamma\gamma$ rate. In fact, from Tab.\ref{tab1}, one may observe that the value of the charged mean radius $\langle r_{\pi^+}^2\rangle^{q\bar{q}}$ runs within the region of $[(0.341 {\rm fm})^2,(0.451 {\rm fm})^2]$ for $B\in[0.00,0.60]$. These values are somewhat smaller than the measured pion charged radius $\langle r^2\rangle^{\pi^+}_{expt} =(0.657\pm 0.012\; {\rm fm})^2$~\cite{radius} and $(0.641{\rm fm})^2$~\cite{jlab}. Since the probability of leading Fock-state $P_{q\bar{q}}$ is less than $1$ and is about $60\%-80\%$, such smaller $\langle r_{\pi^+}^2\rangle^{q\bar{q}}$ for the leading Fock-state WF is reasonable. This confirms the necessity of taking the higher Fock-states into consideration for a sound estimation, especially for small and intermediate $Q^2$ region.

The leading Fock-state pion DA is related with the pion WF through the following relation
\begin{equation}
\phi_\pi(x,\mu_0^2)=\frac{2\sqrt{3}}{f_\pi} \int_{|\mathbf{k}_\perp|^2\leq\mu_0^2} \frac{d^2\mathbf{k}_\perp}{16\pi^3} \Psi_{q\bar{q}}(x,\mathbf{k}_\perp) ,
\end{equation}
where $\mu_0$ stands for some hadronic scale that is of order ${\cal O}(1~{\rm GeV})$. Then, the pion DA takes the following form
\begin{widetext}
\begin{equation}\label{phimodel}
\phi_\pi(x,\mu_0^2) = \frac{\sqrt{3}A m
\beta} {2\sqrt{2}\pi^{3/2}f_\pi} \sqrt{x(1-x)}
\varphi_\pi(x) \left( \mathrm{Erf}
\left[\sqrt{\frac{m^2_q +\mu_0^2}{8\beta^2 x(1-x)}}\right]-
\mathrm{Erf}\left[\sqrt{\frac{m^2_q}{8\beta^2 x(1-x)}}\right]
\right),
\end{equation}
\end{widetext}
where the error function ${\rm Erf}(x)$ is defined as $\mathrm{Erf}(x)=\frac{2} {\sqrt{\pi}} \int_0^x e^{-t^2}dt$. The pion DA at any other scale can be derived through a QCD evolution~\cite{lb,dae}. We call this pion DA-model as the BHL-transverse-momentum improved DA, which has a better end-point behavior and is consistent with the Brodsky and Teramond's holographic model~\cite{bt} that is constructed based on the anti-de Sitter/conformal field theory correspondence.

\begin{figure}
\includegraphics[width=0.35\textwidth]{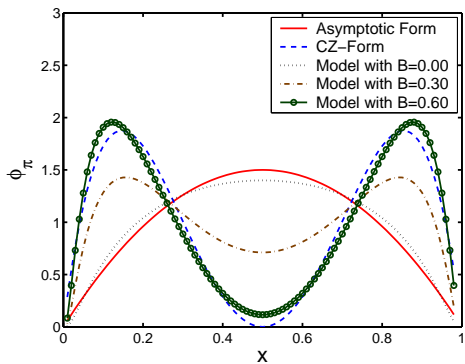}
\caption{Comparison of the pion DA model defined in Eq.(\ref{phimodel}) with the asymptotic-DA and the CZ-DA. } \label{phi}
\end{figure}

The pion DA Gegenbauer moments of $\phi_\pi(x,\mu^2_0)$ can be calculated by the following way \begin{displaymath}
a_n(\mu^2_0)=\frac{\int_0^1 dx \phi_{\pi}(x,\mu^2_0)C^{3/2}_n(2x-1)} {\int_0^1 dx 6x(1-x) [C^{3/2}_n(2x-1)]^2} .
\end{displaymath}
Numerically, it is found that the second Gegenbauer moment $a_2(\mu^2_0)$ is close to the value of $B$ (as shown by Table \ref{tab1}); i.e. the DA's behavior is dominated by $B$ which measures the deviation from the asymptotic form. Moreover, when $B\simeq0.00$, its DA is asymptotic-like; and when $B\simeq0.60$, its DA is CZ-like. This shows $\phi_\pi(x,\mu^2_0)$ can mimic the DA behavior from asymptotic-like to CZ-like naturally by a proper value of $B$. To show this point more clearly, we draw the pion DA by taking $\mu_0=1$ GeV and $m_q=0.30$ GeV in Fig.(\ref{phi}), where $B=0.00$, $0.30$ and $0.60$ respectively.

Next, we do the numerical analysis.

\begin{figure}
\centering
\includegraphics[width=0.4\textwidth]{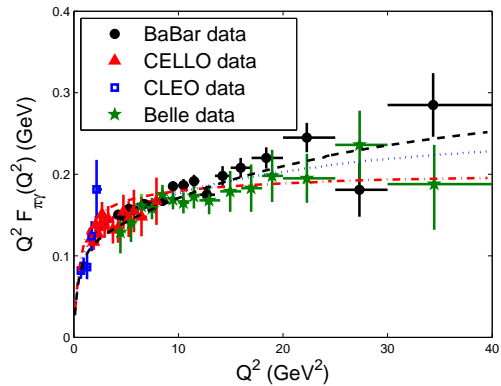}
\caption{$Q^2 F_{\pi\gamma}(Q^2)$ with the model WF (\ref{wave}) by taking $m_q=0.30$ GeV and by varying $B$ within the region of $[0.00,0.60]$. The dash-dot line, the dotted line and the dashed line are for $B=0.00$, $B=0.30$ and $B=0.60$ respectively. } \label{figff1}
\end{figure}

First, we calculate the pion-photon TFF with the model WF (\ref{wave}) by taking $m_q=0.30$ GeV and by varying $B$ within the region of $[0.00,0.60]$. The result is shown in Fig.(\ref{figff1}), where the dash-dot line, the dotted line and the dashed line are for $B=0.00$, $B=0.30$ and $B=0.60$ respectively. The CELLO, CLEO, BABAR and Belle data are included for a comparison. Our present results for $B=0.00$, $B=0.30$ and $B=0.60$ are consistent with three typical predictions for pion-photon TFF derived in the literature, which have been summarized in Ref.\cite{stefanis}. This shows our present pion DA model really provides a convenient model for estimating the pion-photon TFF. In some sense, our present estimation is more reliable, since we require the pion TFF to agree with the more confidently experimental data in small $Q^2$ region simultaneously. For example, it is suggested that a flat pion DA can explain the BABAR's rapid logarithmic-like behavior in large $Q^2$ region of~\cite{flat}, however it fails to explain small $Q^2$ behavior.

In small $Q^2$ region, $Q^2\lesssim 15~GeV^2$, it is found that both the asymptotic-like and the CZ-like DAs can explain the CELLO, CLEO, BABAR and Belle experimental data. Especially, for the CZ-like DAs, because of the suppression from the BHL-transverse-momentum dependence, the end-point contributions have been effectively suppressed, so it can also provide a reasonable estimation of pion-photon TFF. However, at large $Q^2$ region, different DA behavior (by varying $B$) will lead to different pion-photon TFF limiting behavior. Typically, when $Q^2\to\infty$, the $Q^2 F_{\pi \gamma}(Q^2)$ for asymptotic-like DA (with $B=0$) tends to the usual limit $2f_\pi\simeq 0.185 GeV$ \cite{lb}. However to explain the BABAR data on high $Q^2$ region, we need a broader DA with $B\neq0$. With a bigger value of $B$, corresponding to a broader DA as shown by Fig.(\ref{phi}), the estimated pion-photon TFF shall be more close to the BABAR data; while the Belle data prefers asymptotic-like DA with a small $B$. Therefore, the large discrepancy of Belle and BABAR data at the high $Q^2$ region shows we still need more data to determine the pion DA behavior.

\begin{figure}
\centering
\includegraphics[width=0.4\textwidth]{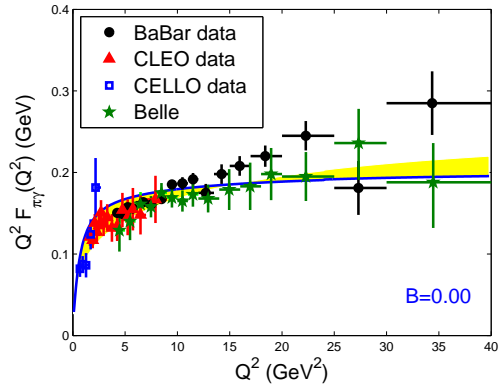}
\caption{$Q^2 F_{\pi\gamma}(Q^2)$ with the model WF (\ref{wave}) by fixing $B=0.00$ (Asymptotic-like DA) and by varying $m_q$ within the region $[0.20,0.30]$ GeV. The solid line is for $m_q=0.30$ GeV, and the shaded band shows its uncertainty.} \label{figffb0}
\end{figure}

\begin{figure}
\centering
\includegraphics[width=0.4\textwidth]{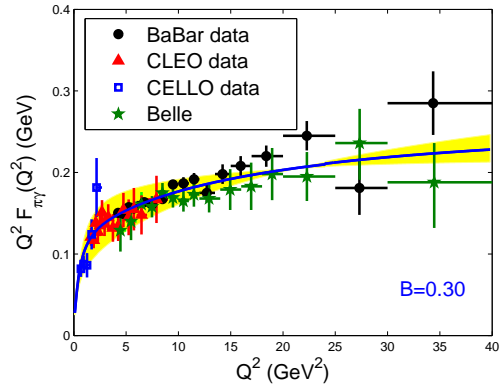}
\caption{$Q^2 F_{\pi\gamma}(Q^2)$ with the model WF (\ref{wave}) by fixing $B=0.30$ and by varying $m_q$ within the region $[0.20,0.40]$ GeV. The solid line is for $m_q=0.30$ GeV, and the shaded band shows its uncertainty.} \label{figffb3}
\end{figure}

\begin{figure}
\centering
\includegraphics[width=0.4\textwidth]{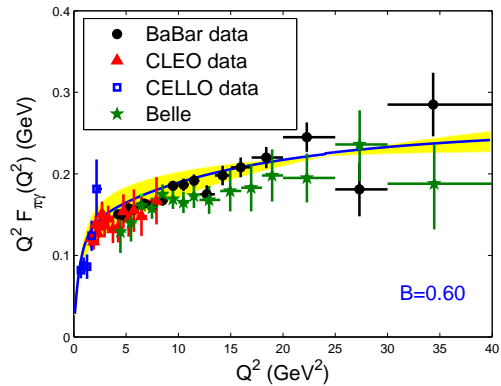}
\caption{$Q^2 F_{\pi\gamma}(Q^2)$ with the model WF (\ref{wave}) by fixing $B=0.60$ (CZ-like DA) and by varying $m_q$ within the region $[0.30,0.50]$ GeV. The solid line is for $m_q=0.40$ GeV, and the shaded band shows its uncertainty.} \label{figffb6}
\end{figure}

Second, we make a discussion on the pion-photon TFF uncertainties by varying the value of $m_q$ under three typical values of $B$, i.e. $B=0.00$, $B=0.30$ and $B=0.60$, respectively. Possible ranges for the DA parameters under different $B$ can be determined by the confidential data at small $Q^2$ region, where the CELLO, CLEO, BABAR and Belle data are consistent with each other. We would like to point that the non-valence-quark term $Q^2 F^{(NV)}_{\pi \gamma}(Q^2)$ gives sizable contributions and should be taken into consideration so as to provide a sound estimation of pion-photon TFF in small $Q^2$ region. Following the same method in Ref.\cite{higher1}, we obtain: $m_q = [0.20,0.30]$ GeV for the case of $B=0.00$; $m_q = [0.20,0.40]$ GeV for the case of $B=0.30$; $m_q =[0.30, 0.50]$ GeV for the case of $B=0.60$ \footnote{A smaller $m_q\lesssim0.20$ GeV will always lead to a probability of the $q\bar{q}$ valence-quark state larger than $1$, so we will not consider it.}. Figs.(\ref{figffb0},\ref{figffb3},\ref{figffb6}) show the pion-photon TFF for $B=0.00$, $0.30$ and $0.60$, respectively. There is a cross-over around $Q_0^2\sim15$ GeV$^2$ for $B=0.00$, $Q_0^2\sim20$ GeV$^2$ for $B=0.30$, $Q^2\sim25$ GeV$^2$ for $B=0.60$; e.g. for the case of $B=0.6$, in the lower $Q^2$ region, the upper edge of the band is for $m_q=0.50$ GeV and the lower edge is for $m_q=0.30$ GeV; while in the higher $Q^2$ region, the upper edge of the band is for $m_q=0.30$ GeV and the lower edge is for $m_q=0.50$ GeV.

In summary, in the present paper, we have recalculated the pion TFF within the light-cone pQCD approach in which both the valence quark state's and the non-valence quark states' contributions have been taken into consideration. For the purpose, we suggest a convenient pionic WF model, whose parameters can be constrained by some physically reasonable constraints and whose DA behavior can be controlled by the parameter $B$. This model can also be adopted for other light pseudoscalar wavefunctions with suitable changes of the constitute quark masses.

In comparison with the present experimental data, our results show that

\begin{itemize}
\item As shown by Figs.(\ref{figffb0},\ref{figffb3},\ref{figffb6}), our estimates for pion-photon TFF by using the pion DA model (\ref{phimodel}) with $B=0.00$, $B=0.30$ and $B=0.60$ accordingly, are consistent with three typical pion-photon TFF predictions derived in the literature, which have been summarized in Ref.\cite{stefanis}. It shows clearly how the asymptotic-like DA and CZ-like DA affect the pion-photon TFF. Then, our present pion DA model provides a convenient way for estimating the pion-photon TFF. Inversely, if we know the pion-photon TFF well, we can conveniently derive the pion DA's correct behavior.

\item Our WF parameters are determined by experimental data at small $Q^2$ region. For $Q^2\lesssim 15~GeV^2$, both asymptotic-like and CZ-like (or even more broader DAs) can explain the CELLO, CLEO, BABAR and Belle experimental data under reasonable choices of WF parameters. In large $Q^2$ region, the new Belle data agrees with the asymptotic DA estimation, while to be consistent with the BABAR data, we need a much broader DA; i.e. the conventional adopted asymptotic DA should be broadened to a certain degree. However the much broader WF/DA will have a serious trouble in producing the correct magnitude of the valence-state structure function of the pion, as pointed out by Ref.\cite{spin1}. Certainly, we believe that it is possible to draw the final conclusion on what the pion DA is, if the more accurate data in the large $Q^2$ region can fix the parameter $B$ in the coming future.

\item Any constructed pseudo-scalar DA models should be consistently explain all the measured pseudo-scalar-photon TFFs, such as $Q^2 F_{\pi\gamma}$, $Q^2 F_{\eta\gamma}$ and $Q^2 F_{\eta'\gamma}$. It has been found that a moderate pseudo-scalar ($\pi$, $\eta$ or $\eta'$) DA with $B\sim 0.1-0.3$ (corresponding to $a_2(\mu_0)\lesssim 0.30$; i.e. close to asymptotic-like behavior) can roughly explain the TFFs $Q^2 F_{\pi\gamma}$, $Q^2 F_{\eta\gamma}$ and $Q^2 F_{\eta'\gamma}$ data simultaneously~\cite{hw2,stan}, especially by introducing a possible amount of intrinsic charm component $f^{c}_{\eta'}$ into $\eta$ and $\eta'$~\cite{hw2,hw5}. such a smaller pion second Gegenbauer moment is consistent with the lattice results  $a_2(1\;{\rm GeV}^2)\sim 0.07$~\cite{a2lattice3}, $a_2(1\;{\rm GeV}^2)\sim 0.38$ \cite{a2lattice1} and $a_2(1\;{\rm GeV}^2)\sim 0.36$~\cite{a2lattice2}.
    
    In this sense, the rapid growth of $Q^2F_{\pi\gamma}$ in high $Q^2$ region observed by BABAR is really amazing. By considering the contributions from higher-twists can not help \cite{stan,agev}. If the BABAR collaboration still insists on their measurements, then there may indeed indicate new physics in these form factors, since it is hard to be explained by the current adopted light-cone pQCD framework.

\item Because of the effective end-point suppression due to the BHL-transverse-momentum dependence, our present model of the pion WF/DA will present a basis for deriving more reliable pQCD estimates. In fact, this BHL-like behavior is helpful for deriving the correct small $Q^2$ behavior. Some applications following the similar idea in constructing the pseudo-scalar meson's twist-3 WF model, providing reasonable power-suppressed twist-3 contributions to the form factors, have already been tried in the literature, c.f. Ref.\cite{twist3}.

\end{itemize}

\hspace{1cm}

{\bf Acknowledgments}: This work was supported in part by Natural Science Foundation of China under Grant No.10975144 and No.11075225, and by the Program for New Century Excellent Talents in University under Grant NO.NCET-10-0882.

\end{document}